\documentclass[preprint]{aastex}

\shorttitle{Star-Formation Driven Biconical Outflows}
\shortauthors{Bizyaev et al.}

\newcommand{\Ha}{H${\alpha}$}

\newcommand{\de}{$^\circ$}

\newcommand{\kms}{km\,s$^{-1}\,$}

\shorttitle{Ionization of eDIG}
\shortauthors{Postnikova and Bizyaev}

\begin{document}

\title{SDSS-IV MaNGA: IONIZATION SOURCES OF DIFFUSE EXTRA-PLANAR GALACTIC MEDIUM}
\author{© 2023 \ \ 
Vera K. Postnikova \altaffilmark{1,2}\footnote{E-mail: vraeranaz@gmail.com },
Dmitry Bizyaev  \altaffilmark{3,1}
}

\altaffiltext{}{\hspace{90pt}$^1$Sternberg Astronomical Institute, Lomonosov Moscow State University, Moscow, Russia}
\altaffiltext{}{\hspace{110pt}$^2$Physics Department of Lomonosov Moscow State University, Moscow, Russia}
\altaffiltext{}{\hspace{90pt}$^3$Apache Point Observatory and New Mexico State University, Sunspot, NM, 88349, USA}

\begin{abstract}
We explore sources of ionization of diffuse gas at different altitudes in galaxies 
in dependence of their stellar mass, \Ha\ luminosity, and specific star formation rate.
We use the MaNGA data from SDSS-IV data release DR16 together with
photoionization and shock ionization models provided by the 3MdB database. 
Our sample comprises 239 edge-on galaxies, which makes our results statistically
valuable. We reach very high galactic altitudes with the help of spectra stacking.
We demonstrate that models of the gas photoionization in a combination of young 
OB-stars and hot low-mass evolved stars (HOLMES) describes the gas ionization 
state in the galaxies of all types on diagnostic diagrams. 
Nevertheless, the shock waves may contribute to the gas ionization in massive
galaxies with passive star formation. We observe a general trend of decreasing 
the fraction of the ionizing flux from OB-stars and the ionization parameter
with the altitude, while the role of the ionization by the HOLMES 
increases. The biggest difference in the contribution from these types of ionizing sources
correlates with the specific star formation rate and with stellar masses
of galaxies. The HOLMES are the principal gas ionization sources in massive galaxies
with passive star formation, while OB-stars dominate the gas ionization in low-mass
galaxies with active star formation.

\end{abstract}


\section*{Introduction}

\citet{hoyle63} started investigating the diffuse ionized gas medium (DIG)
in the Milky Way galaxy, which lead to its detection not only in the galactic midplane,
but also at high galactic altitudes \citep{reynolds73}. Later, the DIG was discovered in 
other galaxies \citep{dettmar90,rand90}. It was found that the DIG phase prevails
at several kpc above the galactic midplane \citep{reynolds91}. A decade ago kinematics of 
the neutral \citep{swaters97, marasco19} and ionized gas \citep{rand00} was quite well studied
only in a few nearby galaxies. Recent progress in massive multi-object spectral extragalactic
surveys enables us to study kinematics of ionized gas in and around objects of the 
local Universe for statistically large samples of galaxies \citep{bizyaev17,levy19,bizyaev22}.

At the same time the ionization sources of the extraplanar gas at large galactic altitudes
(eDIG hereafter) are still not well understood for our and other galaxies. Thus, the ionizing
photons flux from the OB-stars in the galactic midplane is sufficient to explain the amount of 
ionized gas in galaxies with active star formation \citep{haffner09,floresfajardo11}. On the
other hand, the bright forbidden line ratios at high galactic altitudes in some galaxies 
require taking into account evolved stars as the main source of the gas ionization
\citep{floresfajardo11,zhang17,jones17}. The shock wave ionization was also proposed
as a scenario for the explanation of the eDIG emission \citep{collins01}.

A large data release DR16 \citep{ahumada20} of the Mapping Nearby Galaxies at the 
Apache Point Observatory (MaNGA, \citet{bundy15}), a part of the Sloan Digital Sky 
Survey-IV (SDSS-IV, \citet{york00,blanton17}) allows us to make a large sample of 
objects with convenient to observe eDIG. Due to a large number of galaxies, with the 
help of spectra stacking we can trace eDIG to extremely large altitudes above
the galactic midplane — up to a dozen kpc. The main purpose of this study is 
to advance the eDIG studies with the new, large MaNGA sample and new spectra 
modeling results. 

In the next section we describe the MaNGA data that we utilize and their
analysis. Then we describe the used diagnostic diagrams and the line ratio modeling.
Then we report our results and discuss them. Finally, we summarize our results. 
We assume that the Hubble constant is 70 $km\, s^{-1} \, Mpc^{-1}$ throughout our paper. 

\section*{SDSS-IV MaNGA Data} 
\subsection*{The MaNGA Spectra}
We employ data from the MaNGA survey released in the frames of DR16 of SDSS-IV. 
The MaNGA survey was conducted with the 2.5-m Sloan telescope \citep{gunn06}
at the Apache Point Observatory with the resolution of R $\sim$ 2000 and in the 
wavelength range of 3600–10300\AA\ \citep{smee13}. The survey followed 
a sample of over 10,000 galaxies with a uniform distribution by stellar mass 
at the median redshift of z $\approx$ 0.03 \citep{wake17}.

The MaNGA obtained resolved two-dimensional spectral maps
for its objects via two fiber-fed spectrographs \citep{smee13} with 
the Integral Field Unit (IFU) heads \citep{drory15} that consisted of 
densely packed optical fibers allocated at the telescope's focal plane. 
The fiber projection diameter was 2 arcsec, and the spatial filling factor
for the packed circular fibers was 56\%. The full coverage of observed
objects was realized via a 3-point dithering, which allowed to restore
continuous spectra images, see \citet{law15}.

The MaNGA data reduction pipeline consists of two main stages. 
The first stage is the Data Reduction Pipeline (DRP, \citet{law16}), which
delivers flux-calibrated spectra cubes homogenized to the uniform
angular resolution of about 2.5 arcsec (FWHM) placed on a regular 
rectangular spatial grid with a 0.5 arcsec spaxel. The photometric
calibration precision was not worse than 5\% \citep{yan16a,yan16b}.
The second stage is the Data Analysis Pipeline (DAP, \citet{westfall19}),
that separated the absorption and emission spectra using the Penalized
Pixel Fitting (pPXF) method \citep{cappellari04,cappellari17}, which
allowed one to estimate global parameters of galaxies, to obtain
two-dimensional maps of various astrophysical parameters,
cubes of co-added binned spectra, and best-fitting model spectra. 

We make use of the cubes of emission spectra obtained 
after subtracting the model continuum from the observed 
spectra, maps of some emission line fluxes, and gas velocity
maps out of the MaNGA products. We also use some global 
parameters derived from published SDSS photometry. 

\subsection*{Making Masks for Selected Galaxies}
 
We create spaxel masks to select only good quality data for the consequent
spectra stacking. We leave only spaxels that satisfy the following
criteria:

\begin{itemize}

\item the emission spectrum was successfully modeled 
for this spaxel, with no bad data processing flags;

\item the radial velocity in the \Ha\ emission line was successfully determined 
for this spaxel, with no bad data processing flags;

\item the signal-to-noise ratio (SNR) in the \Ha\ line was $\ge$ 3.0;

\item the absolute value of the radial velocity for this spaxel was within 
350 \kms of the galactic center velocity. 

\end{itemize}

\subsection*{Analysis of the Sample}
 
The primary selection of edge-on galaxies was performed via the visual inspection 
of composite, color images made by SDSS survey. Our experience in selecting
edge-on objects for large catalogs \citep{bizyaev14} and for individual studies
of MaNGA objects \citep{bizyaev17} shows that the visibility of the dust lane projected to
the central region of galaxies suggests high inclinations of galactic midplane
to the line of sight $\ge$ 85\de, which also has been confirmed 
with calculations by \citet{bizyaev04,mosenkov15}. 
In turn, the high inclination ensures that we can study highly elevated gas
without its overlapping on bright star formation regions in the midplanes of galaxies. 
Our result of the visual selection is a sample of 258 edge-on galaxies. 

After the visual inspection we notice that some galaxies should be rejected based
on their maps of the \Ha\ emission, maps of the equivalent width EW(\Ha), and 
maps of the stellar and gas velocities. Thus, our sample has a few objects in
which their gas and stars rotate in near orthogonal planes, which resembles
galaxies with polar rings, see e.\,g. \citet{moiseev11}. While they can be interesting 
objects for the further studies, they don't allow us to study eDIG by the methods that 
this paper uses. As a result, we reject 32 more
galaxies that have one or more of the following features:

\begin{itemize}

\item a large angle between the stellar and gas rotation;

\item evident inconsistency between reported photometric parameters 
and observed picture, e.\,g. when the effective radius $R_{eff}$ is
too large with respect to the visible size of the object;

\item the spaxel mask rejects the majority of galactic regions,
which can statistically bias the contribution of the remained 
spaxels from this object to the stacked spectra.

The resulting sample comprises 239 galaxies. We show them as a mosaic in Figure 1. 

\end{itemize} 

\begin{figure}
\epsscale{1.00}
\plotone{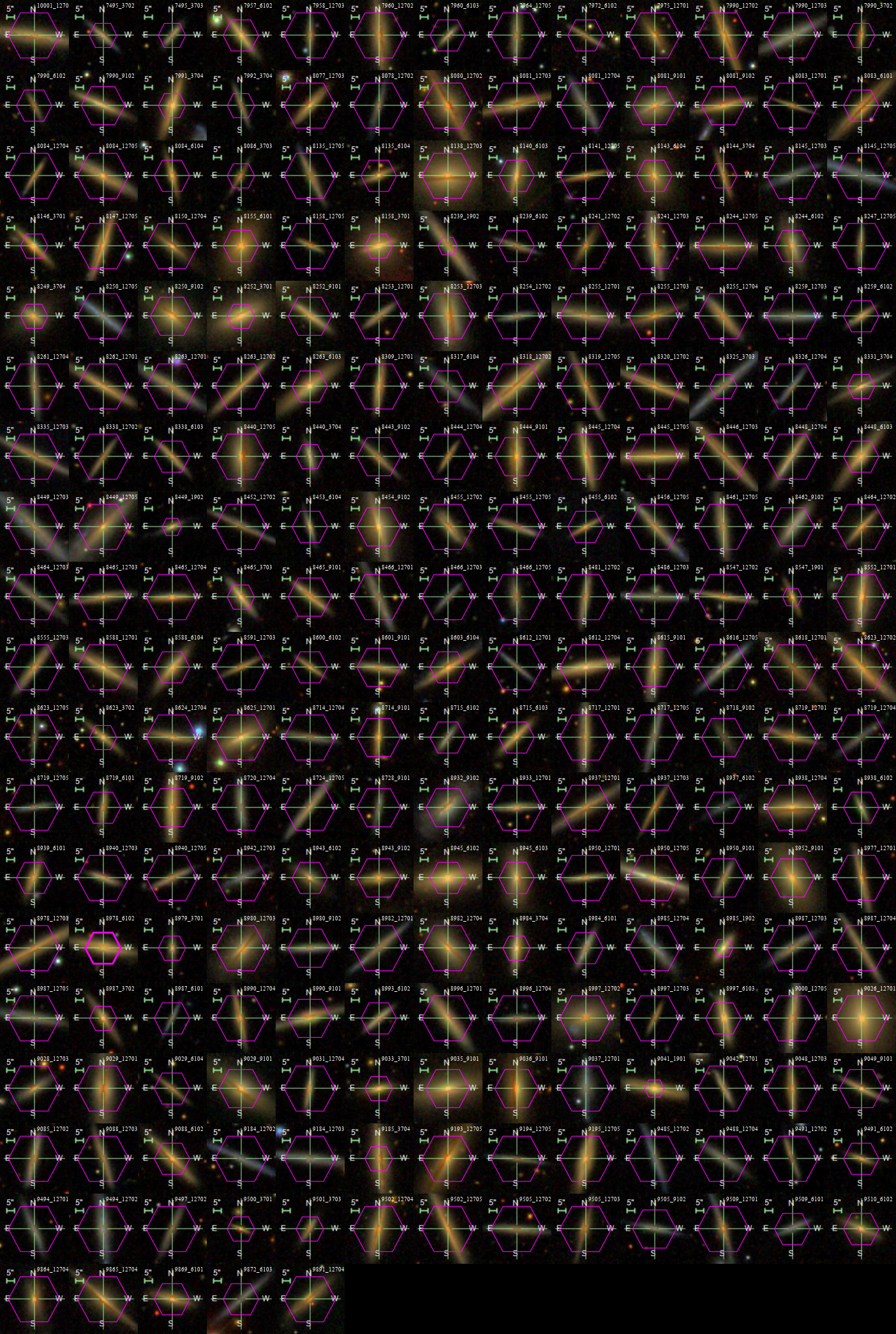}
\caption{\small The mosaic of 239 galaxies from our sample. The magenta hexagons that designate
the MaNGA IFU coverage overlap on SDSS color images. 
\label{fig:image_1.png}}
\end{figure}

\subsection*{The Spectra Stacking Procedure and the Emission Line Flux Estimation}
 
Since we study faint regions far away from galactic midplanes, the individual SNR
of their emission lines is often too low for the analysis. To increase the SNR, we stack 
spectra from regions that are close by their properties. To do it, we subdivide the sample by 
a small number of groups, or bins, with relatively similar global parameters.
Among with studying the eDIG properties at different altitude bins, we also 
incorporate a binning by the following galactic parameters:

\begin{itemize}

\item the integral stellar mass $M_s$ estimated by the  NASA-Sloan Atlas of galaxies 
(NSA\footnote{\url{http://nsatlas.org}});

\item the galactic luminosity not corrected for the reddening $L_{H\alpha-R_{eff}(r)}$,
which is the \Ha\ luminosity within one effective radius $R_{eff}$ in the $r$-band, and proportional to the star formation rate (SFR), and also taken
from the NSA; 

\item the specific star formation rate $sSFR$ estimated as 
$sSFR = L_{H\alpha-R_{eff}(r)} / 10^{41.27} / M_s$, 
where the normalization is derived by \citet{kennicutt12}, \citet{murphy11}, \citet{hao11};  

\item the visual altitude of spaxels $z/z_0$ above the galactic midplane normalized by 
the exponential scale height, the latter was estimated as $z_0=0.596 \cdot R_{eff} \cdot b/a$, 
where $R_{eff}$ is the effective radius, and $b/a$ is the minor-to-major axis ratio 
from a 2D fitting in the $r$-band, both taken from the NSA estimated in the $r$-band with the Petrosian parameterization. 

\end{itemize}

We optimize the binning via selecting the number of bins and their borders
in the way that in each bin by the galactic parameters and by the altitude
has a contribution from at least 10\% of all galaxies in the sample. After that we co-add corrected for the radial velocities emission spectra in each bin. We ensure that the corrected spectra have no other lines 
within $\pm$ 7.5\AA\ from the line of interest. We also check and confirm that 
the DAP subtracted continuum spectra so well that an additional continuum
correction is not required. The flux in selected emission lines is
found via a simple integration of the line intensity profiles within $\pm$ 7.5\AA\
from the line centers. Then we correct the line fluxes for the 
extinction based on the Balmer decrement. Since the DAP provides uncertainties
of the emission intensities in each spaxel, we find the resulting flux
uncertainties via the standard error propagation method.

As a result, we co-add the emission spectra for 239 galaxies in the optimal 
bins by the general galactic parameters and by the galactic altitude
according to the procedure described above. Then we find the emission line
fluxes and their uncertainties. An example of our binning by the galactic altitudes
is shown in Figure 2.


\begin{figure}
\begin{minipage}[h]{0.45\textwidth}
\center{\includegraphics[scale=0.31]{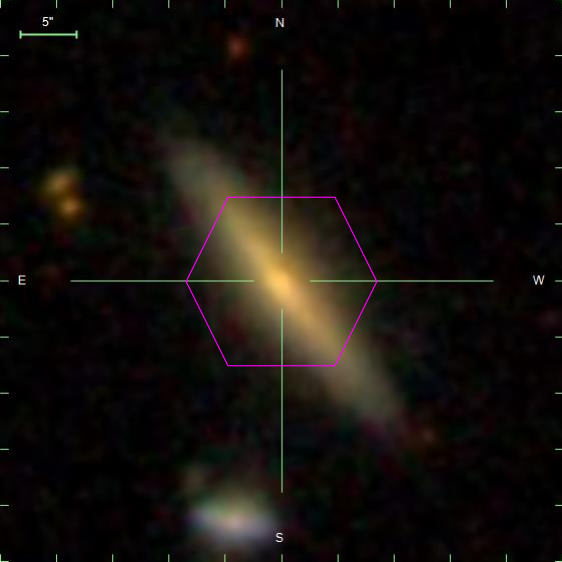}}
\end{minipage}
\hfill
\begin{minipage}[h]{0.55\textwidth}
\center{\includegraphics[scale=0.65]{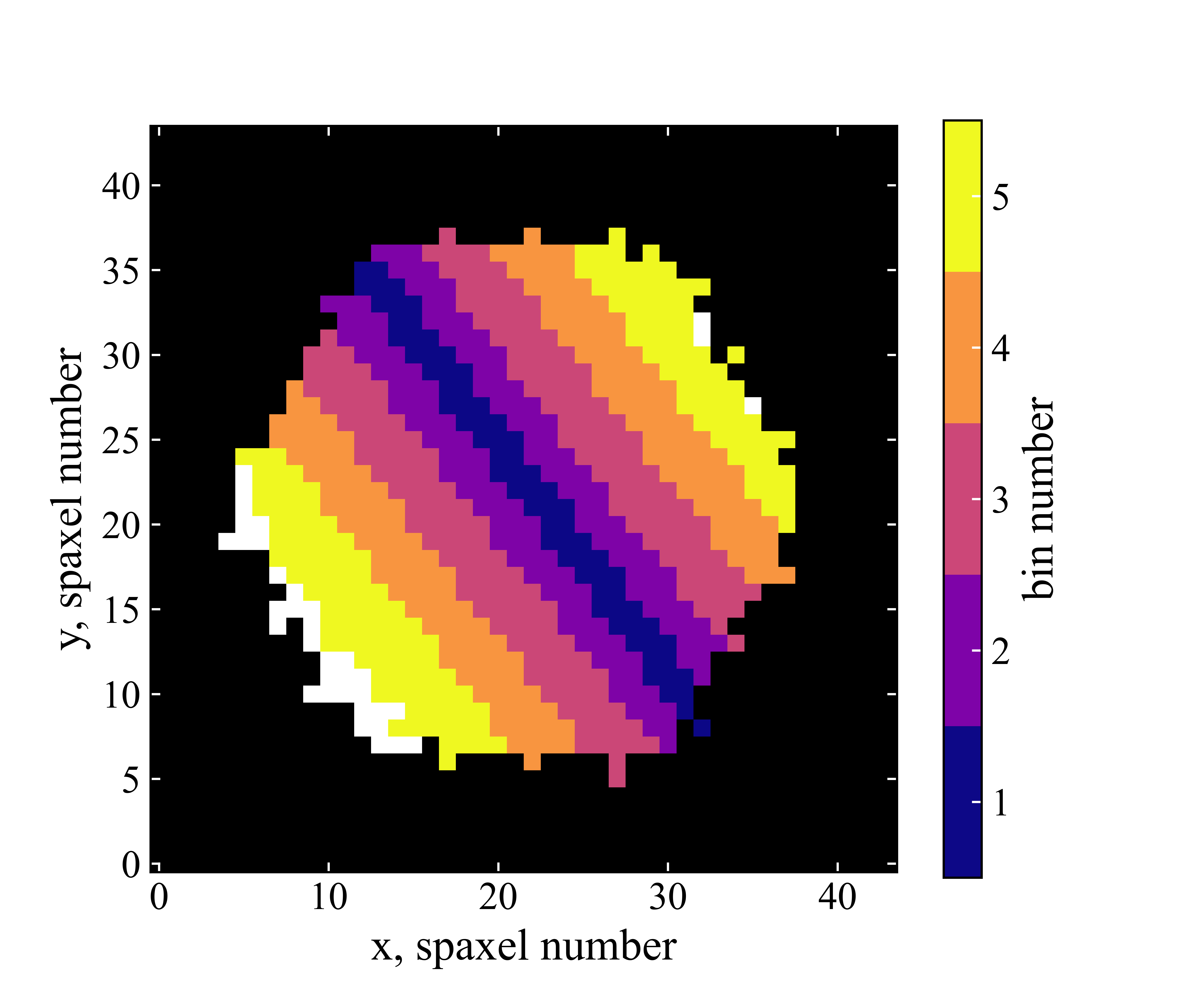}}
\end{minipage}
\caption{\small Left: an SDSS image of one of our galaxies with superimposed IFU area outlined by 
the magenta hexagon. Right: corresponding bins by the galactic altitude with the spaxel mask shown by the white color. 
\label{ris:gal_bin_im.png}}
\end{figure}

\section*{Diagnostic Diagrams and Theoretical Models}
\subsection*{Diagnostic Diagrams}
 
In order to compare the data of observations with models, we
employ diagnostic diagrams that are based on relative intensities 
of emission lines. They enable us to efficiently separate regions with different 
physical conditions. The diagnostic diagrams are widely
used after a study by \citet{bpt}, where the advantage of 
two-dimensional classification was demonstrated and also the most
useful combinations of strong emission lines were introduced. 
\citet{veilleux87} also considered a set of the line ratios
on diagnostic diagrams, and both works contributed to a set 
of the diagrams that are traditionally called as BPT-diagrams. \citet{dopita00}
explained and updated the classification with the help of models
that described the gaseous medium. 

An important addition to the diagnostic diagrams are the demarcation lines, 
which show borders between the gas with different ionization mechanisms. 
The most frequently used demarcation lines were introduced by 
\citet{kewley01}, \citet{kauffmann03}, \citet{kewley06}. A more advanced approach to the 
demarcation line setting that utilizes gas dynamics was considered
by \citet{law21}. The data of MaNGA allow us to use a variety of 
emission line combinations for the diagnostic diagrams, but in this 
work we limit the study by 3 traditional BPT diagrams:

\begin{itemize}
	\item log([OIII]$\lambda$5007/H$\beta$) vs log([NII]$\lambda$6583/H$\alpha$);
	\item log([OIII]$\lambda$5007/H$\beta$) vs log([SII]$\lambda\lambda$6716,6731/H$\alpha$);
	\item log([OIII]$\lambda$5007/H$\beta$) vs log([OI]$\lambda$6300/H$\alpha$).
\end{itemize}

\subsection*{The 3MdB Models}
 
The models of the gas ionization based on numerical computations
are widely demanded for studies of interstellar medium. At the same time,
running the computations is a time-consuming process, and they are 
run and published mostly for limited and specific cases. This problem was mitigated by 
the creation of the Mexical Million Models dataBase (3MdB) by 
\citet{morisset15} and \citet{alarie19}. The 3MdB contains 
pre-computed model emission line ratios that are organized as a
MySQL database. This approach allows one to save time and computation
resources and simplifies comparison between observations
and modeling. 

At the moment the 3MdB consists of two major parts:

\begin{itemize}
\item The database of photoionization models 3MdB-p \citep{morisset15} computed with 
the package Cloudy \citep{ferland98}, version C13 \citep{ferland13}.
This database considers several principal setups of the photoionization modeling
and includes the grids ''DIG\_HR'' — the models designed for the DIG (and eDIG) 
description. They assume that the gas is ionized by a combination of 
radiation from OB-stars and from HOt Low-Mass Evolved Stars (HOLMES).
The choice of these two main sources of the gas ionization is based on 
a study of a well-known edge-on galaxy NGC~891 \citep{floresfajardo11}.
One of the purposes of our study is to verify whether the generalization 
of this assumption is suitable for all types of galaxies.

\item The database of shock ionization models 3MdB-s \citep{alarie19} computed
with the help of the MAPPINGS astrophysical plasma modeling 
code \citep{binette85}, version V \citep{sutherland18}.
Out of cases considered by this set of models, the most relevant are
the models ''Allen08'' described by \citet{allen08}.
Another goal that we pursue in this paper is to verify whether 
the shock models can be relevant to the ionization of gas at different
galactic altitudes in various types of galaxies.  

\end{itemize}

\subsection*{The Photoionization models ''DIG\_HR''}
 
The photoionization grids, that describe the diffuse gas in the frames
of the 3MdB-p database, have 4 model parameters. 

\begin{enumerate}

\item The flux from the OB-stars $\Phi_{OB}, \; photons/sec/cm^2$.
It spans the range $\log\,\Phi_{OB}=(3.5 \div 7.5)$
with the step of 0.25 dex.

\item The ionization parameter $U=\Phi_{total}/n_e/c$, where $\Phi_{total}=\Phi_{OB}+ \Phi_{HOLMES}$, $n_e$ is the electron density, and $c$ is the speed of light. 
It spans the range $\log\,U=(-4.0 \div -3.0)$ with the step of 0.1 dex.

\item The gas metallicity $O/H$ defined as the ratio of the number of corresponding atoms. 
It spans the range $\Delta\,O/H= (-1.0 \div 0.6)$ 
with the step of 0.1 dex, where $\Delta\,O/H + 8.69 = 12.00 + \log\,O/H$.

\item The nitrogen abundance $N/O$ defined as the ratio of the number of corresponding atoms.
It spans the range $\log\,N/O=(-1.4 \div -0.2)$
with the step of 0.1 dex.

\end{enumerate}

It is important to notice that the HOLMES flux is fixed 
at $\Phi_{HOLMES}= 8.4 \cdot 10^4 \; photons/sec/cm^2$ in all ''DIG\_HR''
models. 
All other metal abundances relative to O are fixed to the solar values except for the Mg, Si, 
and Fe, which are decreased by 1 dex with respect
to the solar values. 

\subsection*{The shock models ''Allen08''}
 
The grids of models for the shock ionization ''Allen08''
have also 4 model parameters:

\begin{enumerate}

\item Pre-shock gas density. We consider only two realistic 
in our case values: $n=0.1 \; cm^{-3}$ and $n=0.01 \; cm^{-3}$. 

\item The gas metallicity. We have to stick to the solar metallicities
for these grids because the lower metallicity cases, e.\,g. typical for
the Large and Small Magellanic Clouds, are not computed for very
low gas densities typical for the eDIG. We analyze possible behavior of 
the models for the eDIG set of parameters below.  

\item The magnetic field. In the midplane of our and other galaxies
the magnetic field has $B \simeq 10 \; \mu G$, which decreases with the altitude
down to $B \simeq 5 \; \mu G$ at several kpc above the midplane. We consider a 
range of the $B$ between \mbox{1 and 10 $\mu G$}. Within this range the ''Allen08'' grids 
are computed for the 
$B=(1.0,\:1.26,\:1.58,\:2.0,\:3.16,\:4.0,\:5.0,\:10.0) \; \mu G$ with 
$n=0.1 \; cm^{-3}$ and for the $B=(1.0,\:10.0) \; \mu G$ with $n=0.01 \; cm^{-3}$.

\item The shock wave velocity. We consider all provided values
of this parameter. 
For the grids with $n=0.1 \; cm^{-3}$ the velocities range between 
100 and 1000 \kms, and with $n=0.01 \; cm^{-3}$ they range between
200 and 1000 \kms, both with the step of 25 \kms.  

\end{enumerate}

\section*{Results}

\subsection*{The optimality of our binning}
 
Our binning scheme fills all altitude bins with statistically valuable number
of  representatives. When only the altitude binning is considered, 
about 30\% of all galaxies contribute to the highest, least populated bin. 
When an additional galaxy parameter binning is added, some 10\% of all galaxies 
have representative data in the highest bin. 

We ensure that our binning scheme is stable against the number of galaxies in
the parental sample, the number of altitude bins, and the bin altitude
border values by changing these values and inspecting the resulting diagrams
explained below. 

\subsection*{Co-added Galactic Spectra in the Diagnostic Diagrams}
 
Here we consider how the line ratios of our co-added spectra
overlap with the model photoionization and shock grids in dependence of 
our integral galactic parameters $M_s$, $L_{H\alpha-R_{eff}(r)}$, and $sSFR$.
Note that we vary the $\Delta\,O/H$ and $\log\,N/O$ for the
photoionization models in dependence on galactic parameters, which is 
discussed below. The results are shown in Figures 3, 4, and 5. 
 
\begin{figure}
\epsscale{1.00}
\plotone{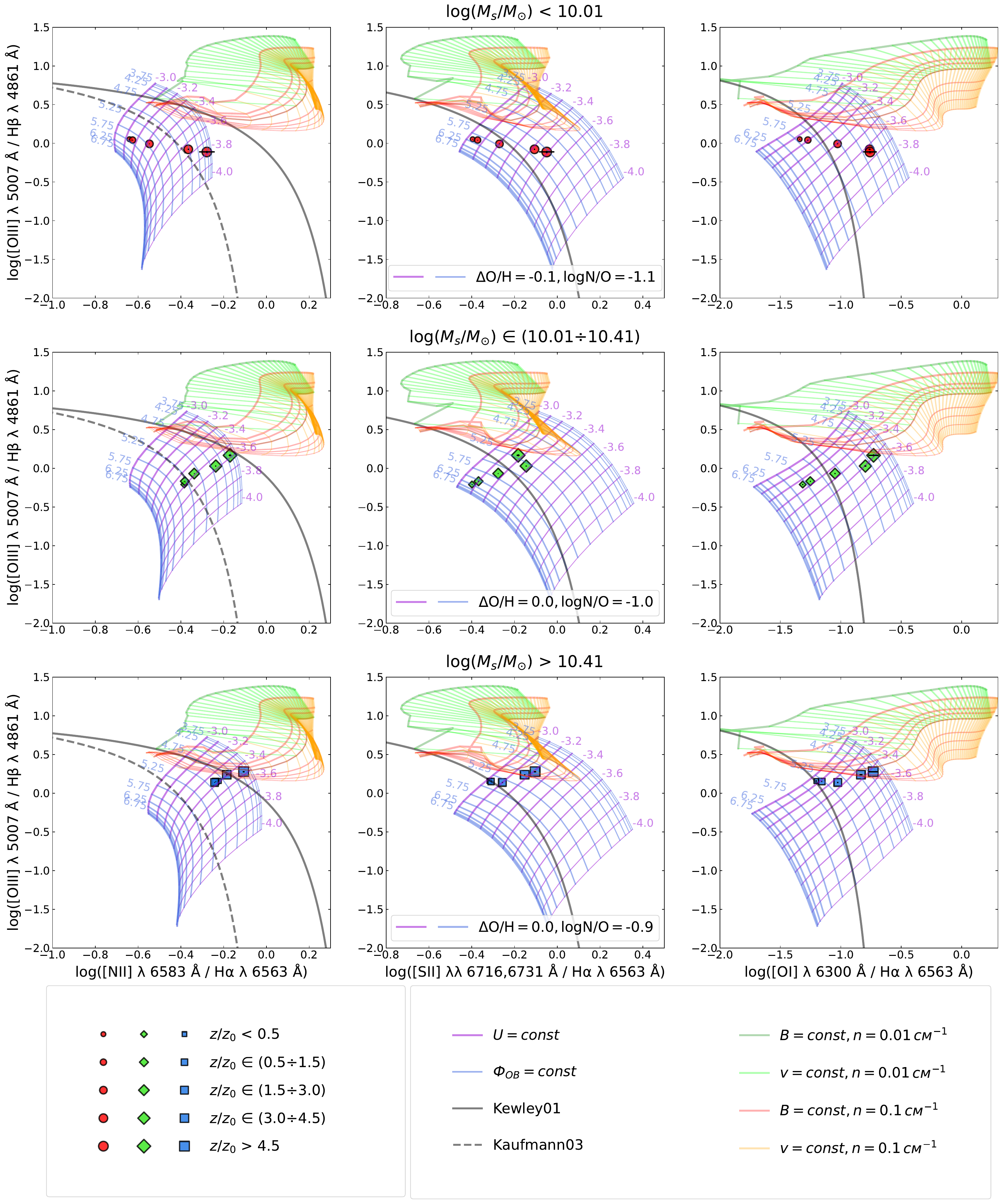}
\caption{\small BPT-diagrams with our observing flux ratios shown for different altitude 
and stellar mass $M_s$ bins. The bigger symbol size corresponds to a higher altitude. The red
circles, green rhombuses and blue squares designate the lower, intermediate, and higher bins
in the $M_s$, respectively. 
The model photoionization (blue) and shock (red and green) grids (see text) are
also shown. The thicker lines designate the greater values of the corresponding parameter. Since we are focused on the former models, the shock grids are 
shown for comparison purposes and their parameters are not specified in the plot.
The demarcation lines from \citet{kewley01} and \citet{kauffmann03} are plotted 
with the solid and dashed curves, respectively. 
\label{fig:image_3_eng.pdf}  }
\end{figure}

\begin{figure}
\epsscale{1.00}
\plotone{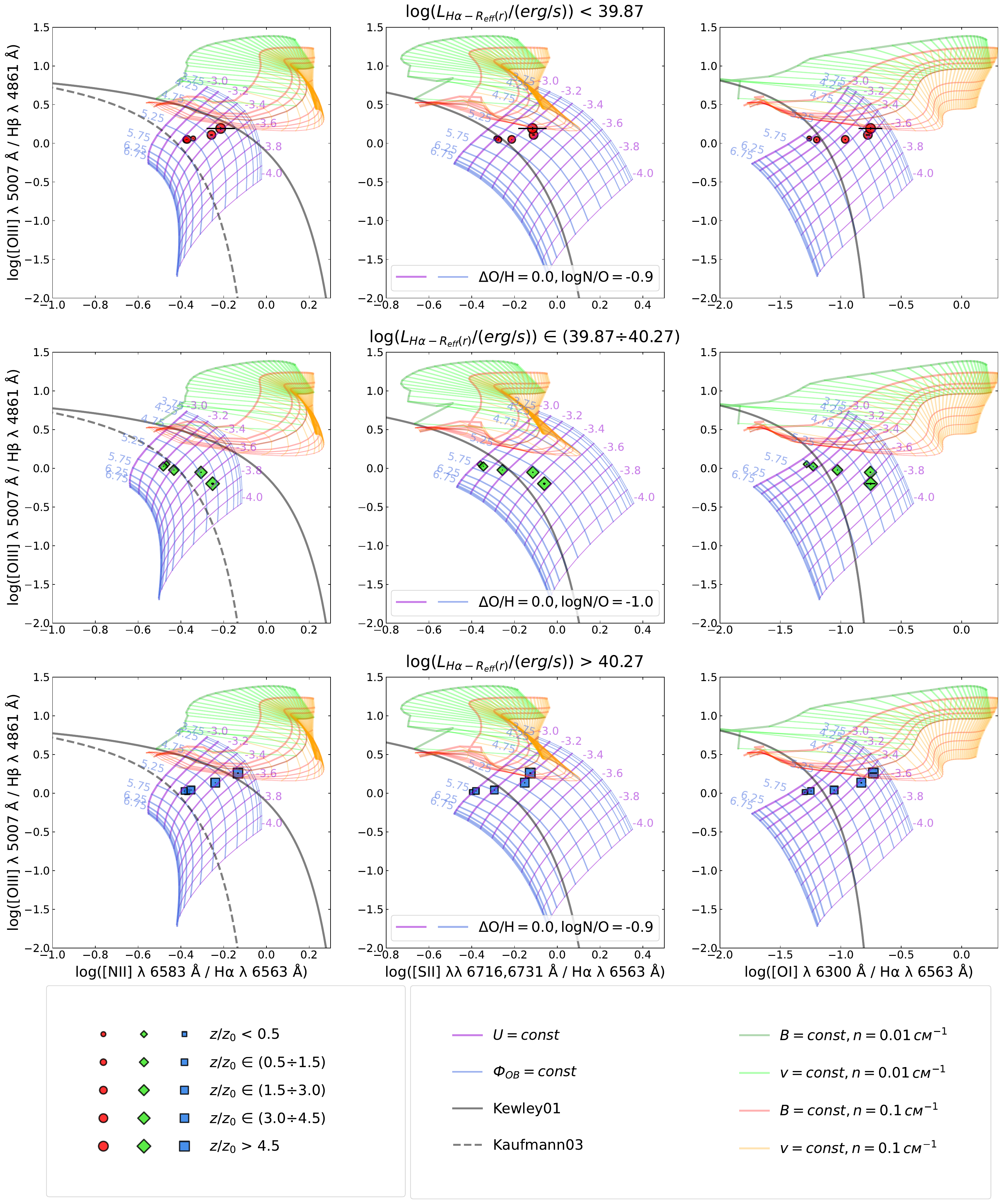}
\caption{\small BPT-diagrams with our observing flux ratios shown for different altitude 
and $L_{H\alpha-R_{eff}(r)}$ bins. All designations are kept the same as in Figure 3.
\label{fig:image_4_eng.pdf} }
\end{figure}

\begin{figure}
\epsscale{1.00}
\plotone{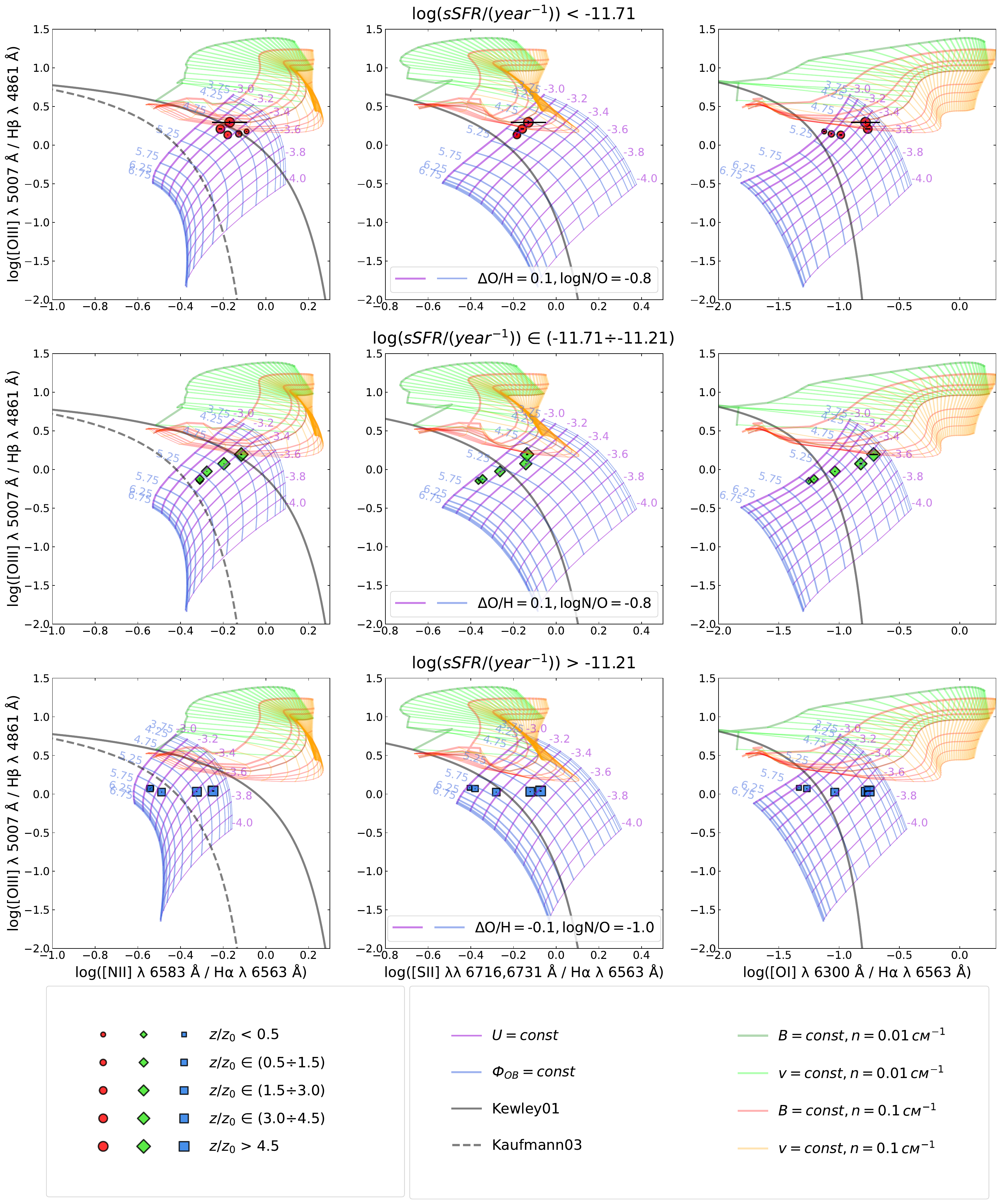}
\caption{\small BPT-diagrams with our observing flux ratios shown for different altitude 
and $sSFR$ bins. All designations are kept the same as in Figure 3.
\label{fig:image_5_eng.pdf} }
\end{figure}

\subsection*{The Data Interpolation on the Diagnostic Diagrams}
 
As we can see in Figures 3–5, some high-altitude data for the intermediate and high $M_s$, 
for the low and high  $L_{H\alpha-R_{eff}(r)}$ and for low and intermediate $sSFR$
fall onto the cross-section of the photoionization and shock grids. Nevertheless,
all our observing line ratios can be explained with the photoionization
grids only. The position of the points on the BPT diagrams can be 
translated to the OB-stars relative flux and the ionization parameter
via a grid interpolation procedure. Below we estimate these parameters 
by neglecting the shocks contribution to the gas ionization.

At the beginning, let's consider our simplest binning scheme by galactic altitudes only. 
We note that the $O/H$ ratio should change with the altitude, but using photoionization
models for HII regions from \citet{dopita16} we conclude that the $O/H$ variation
with the altitude can be neglected in the frames of the calibration uncertainties 
of $\pm$0.1 dex. When different BPT diagrams are compared between each other, 
the best agreement for our data can be achieved for $\Delta\,O/H = 0.0$ and $\log\,N/O = -0.9$, 
which corresponds to the solar abundance. 
Note that the models do not allow us to vary the S abundance, which leads to systematically
bias of all diagrams that include the S. To avoid the bias, for the regression below 
we utilize only the BPT-plots based on H, N, and O. We also exclude the very first, 
in-midplane bin from the further regression because a very high dust extinction there can 
bias line ratios even for nearby lines in spectra. The results of the grid interpolation and 
regression are shown in Figure 6.

\begin{figure}
\epsscale{1.00}
\plotone{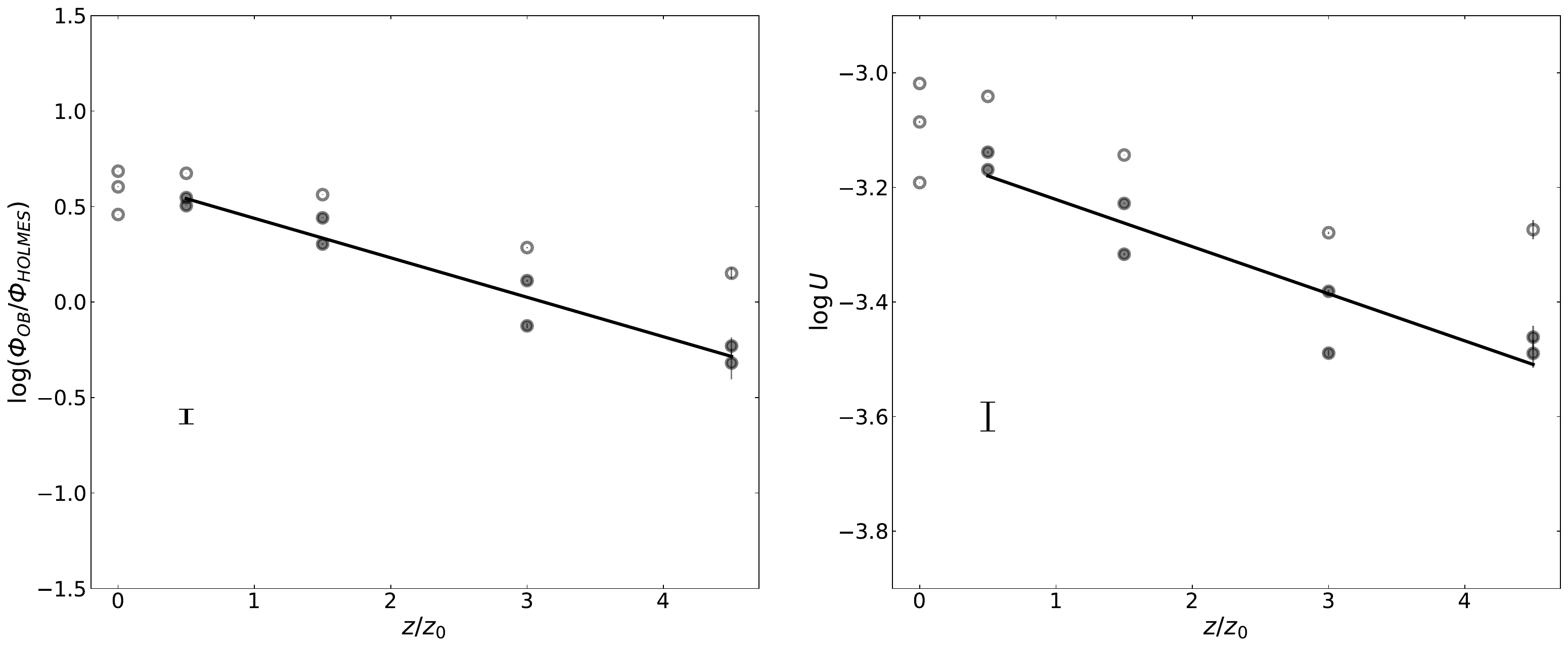}
\caption{\small Results of the grid interpolation for the case of a general binning 
of all galaxies by the galactic altitude without involving any other galactic 
parameters. The ratio of OB-stars and HOLMES fluxes (left) and the ionization
parameter (right) are shown with symbols, and the linear regression 
results are shown with lines. The error bars correspond to one sigma
standard deviation. Open symbols are excluded from the regression (see text). 
\label{fig:image_6_eng.pdf} }
\end{figure}

Next, we consider more complicated binning cases and include additional galactic
parameters introduced above. Before the interpolation, we also use results of HII
regions modeling from \citet{dopita16} and ensure that the $O/H$
ratio would not change by more than $\pm$0.1 dex in the case of the diagrams
for case of adding the $M_s$, $L_{H\alpha-R_{eff}(r)}$, or $sSFR$ to the binning
schemes, hence we conclude that we can neglect the $O/H$ variation with the galactic altitude. 
Then we make the plots similar to Figure 6 that also include additional galactic
parameters and find $\Delta\,O/H$ and $\log\,N/O$ for which the extracted 
parameters best agree between all our BPT plots. 

We notice that the best values of $\Delta\,O/H$ and $\log\,N/O$ for the 
galaxies with different $M_s$ (Figure 3) correlate with $M_s$ same way
as in well-known $O/H$ – $M_s$ \citep[see e.\,g.][]{tremonti04,andrews13,curti17},
$N/O$ – $M_s$ \citep{perezmontero09,andrews13,masters16}, and $N/O$ – $O/H$
\citep{andrews13} relations. 

As for Figure 6, we derive the best-fitting model parameters 
for each observing bin by interpolating
among the model grids on the BPT diagrams. Then for the regression we include the BPT diagrams with H, N, and O only, and  exclude the lowest galactic altitude points. The altitude distribution of the extracted 
parameters and the linear regression lines are shown in Figures 7–9.

\begin{figure}
\epsscale{1.0}
\plotone{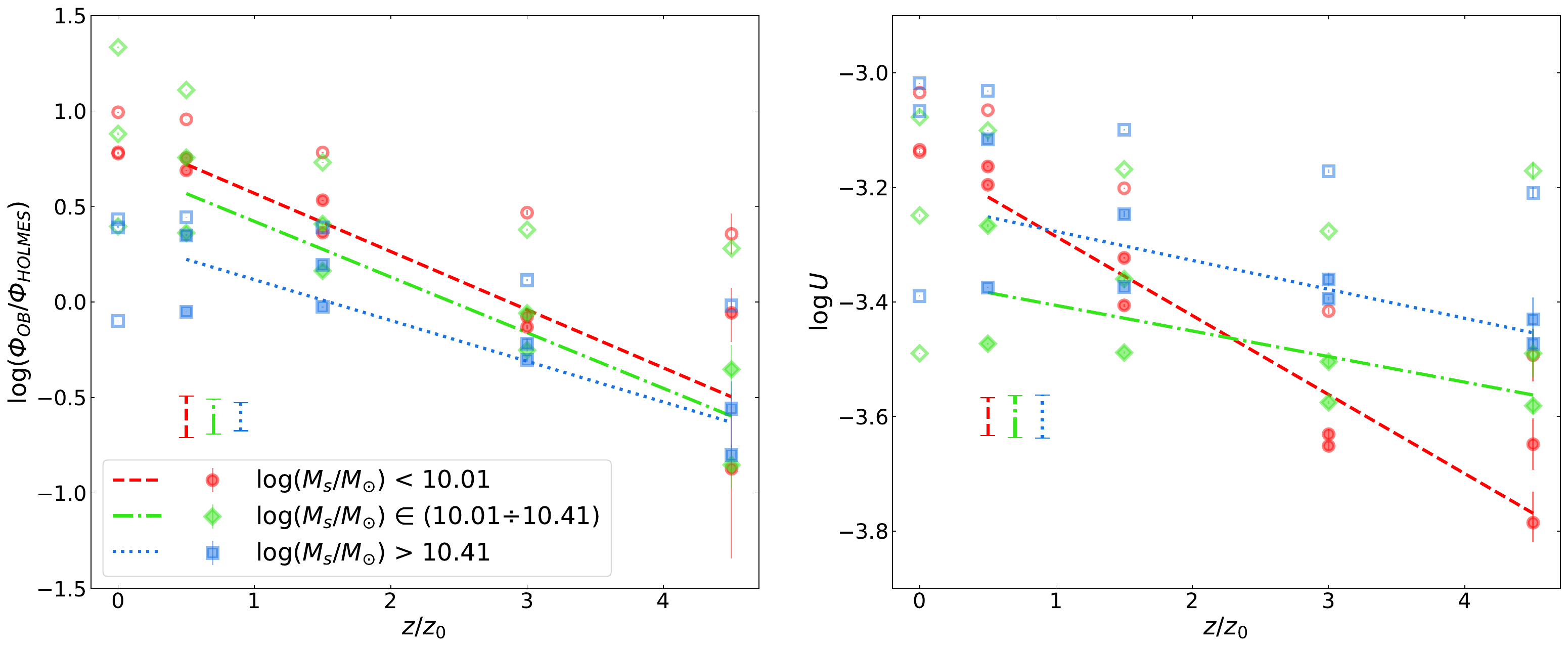}
\caption{\small The altitudinal distributions of the extracted parameters, similar
to Figure 6, but in this case the binning scheme includes stellar masses  $M_s$.
The red
circles with dashed line, green rhombuses with dot-dashed line, and blue squares with dotted
line designate the lower, intermediate, and higher bins
in the $M_s$, respectively. 
\label{fig:image_7_eng.pdf}  }
\end{figure}

\begin{figure}
\epsscale{1.0}
\plotone{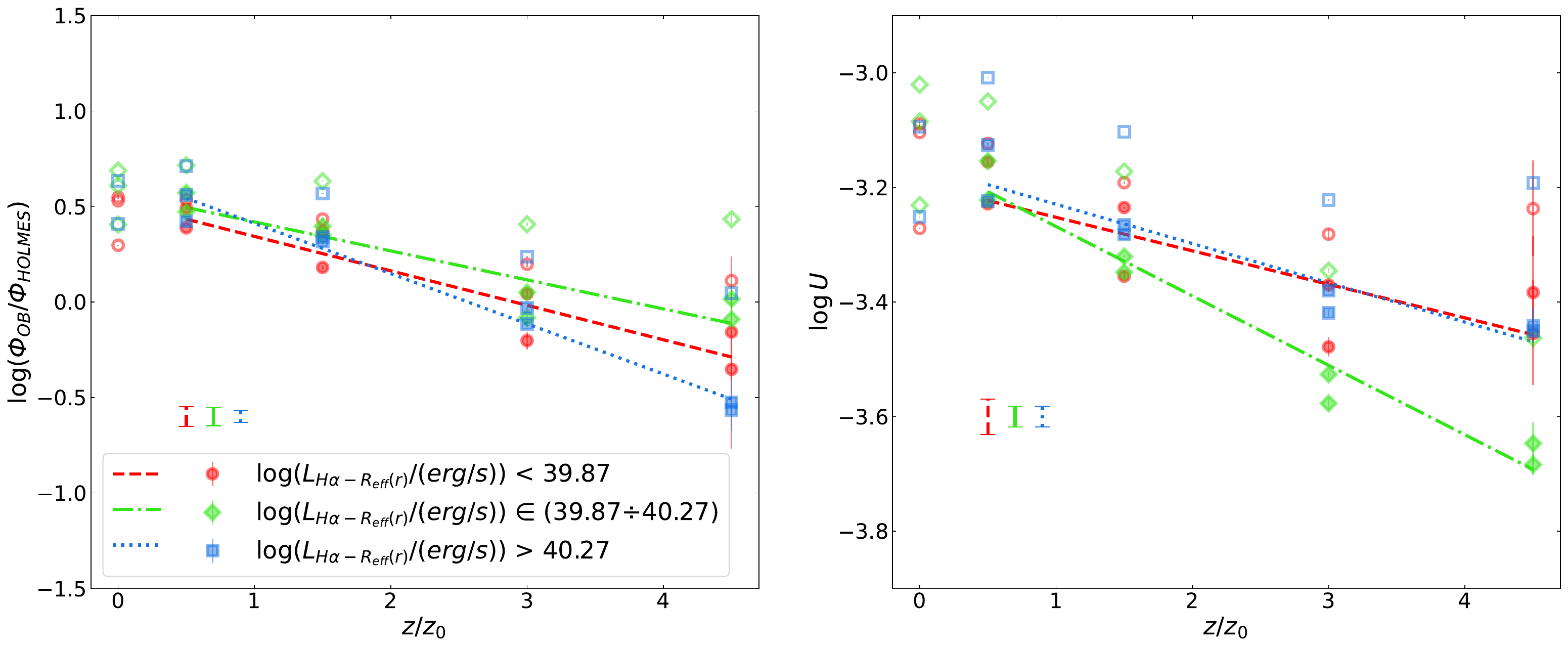}
\caption{\small Same as Figure 7 for the case of binning by the \Ha\ luminosity
within one $R_{eff}$ in the $r$-band, $L_{H\alpha-R_{eff}(r)}$.
\label{fig:image_8_eng.pdf} }
\end{figure}

\begin{figure}
\epsscale{1.0}
\plotone{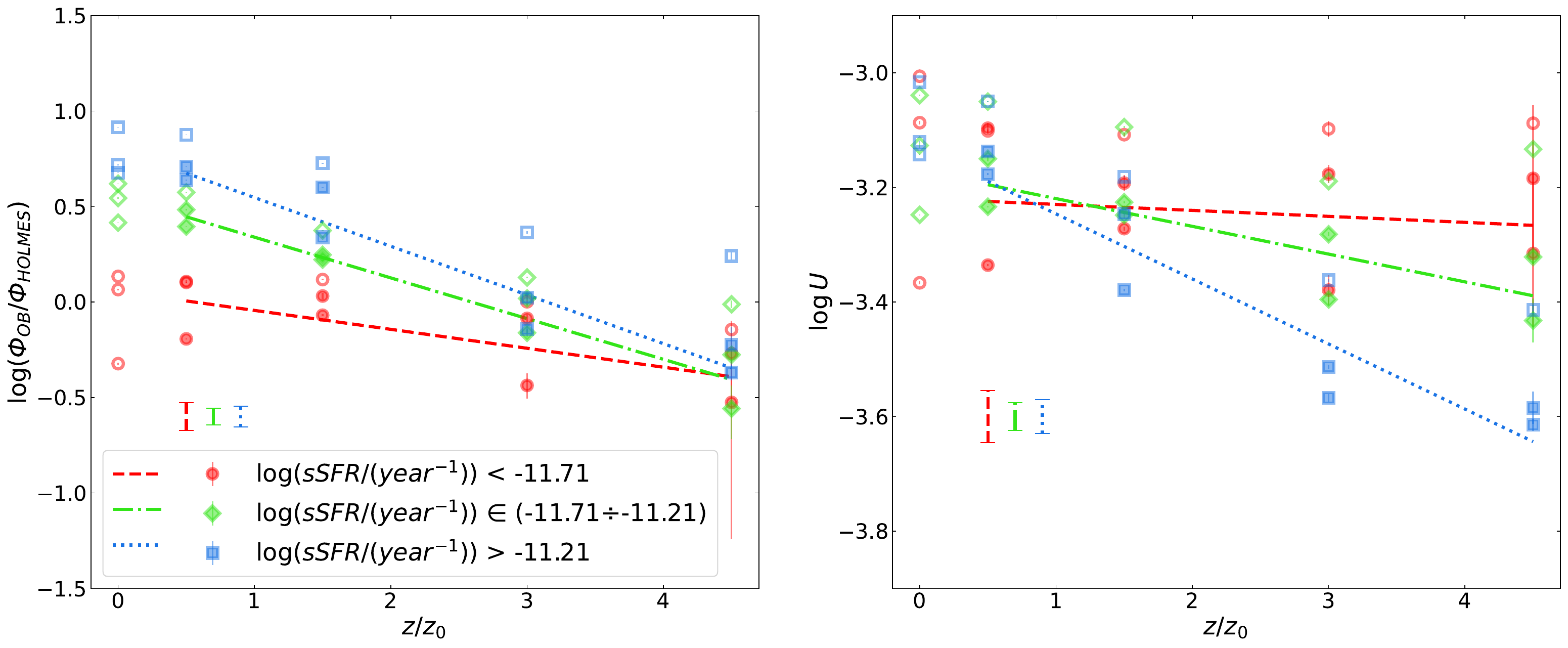}
\caption{\small Same as Figure 7 for the case of binning by the $sSFR$.
\label{fig:image_9_eng.pdf} }
\end{figure}

\section*{Discussion} 
 
Before making certain conclusion about possible contribution of shocks to the DIG ionization we
need to verify how would the shock grids move in Figures 3–5 for the lower than solar
gas metallicities, which is expected for eDIG at high galactic altitudes. As we have
noticed above, grids for subsolar metallicities were computed in the 3MdB only 
with $n= 1 \; cm^{-3}$, which is way higher than typical eDIG densities. Therefore, 
we consider the direction of trends for $n= 1 \; cm^{-3}$ and assume that the 
trends will be the same for lower gas densities. The gas metallicities can be ordered
as (1) twice as solar, (2) solar, (3) subsolar with the Large Magellanic Cloud
values, and (4) subsolar with the Small Magellanic Cloud values. Comparing the model
grids in this order,
one can see that the shock grids move from the right to the left
in Figures 3–5, parallel to the abscissa axes. In this case the position of the 
shock grids on Figures 3–5 will not overlap more with the observing points and the 
conclusions are expected to remain the same for all low gas metallicities. 

The eDIG emission line ratios on the BPT diagrams can be explained with 
two sources of ionization only: hot massive stars and HOLMES. This is 
important to stress that OB-stars alone cannot explain the line ratios 
in the eDIG at high altitudes, and incorporating HOLMES is required. 

Nevertheless, we notice that some sets of parameters, like 
$z > 4.5 \, z_0$ for high and intermediate masses $M_s$ (Figure 3),
lowest and highest $L_{H\alpha-R_{eff}(r)}$ (Figure 4), or 
lowest and intermediate $sSFR$ (Figure 5) send observing
points to the areas where the photoionization and shock grids 
overlap. This may suggest a non-negligeable contribution
of shock waves to the eDIG ionization in those types of galaxies. 

As we remind above, the assumption of two types of sources (OB-stars and HOLMES) 
that ionize gas in the ''DIG\_HR'' photoionization models was based on the study
by \citet{floresfajardo11} that relies on a single galaxy, NGC~891, in which
the gas was traced up to 4 kpc above the galactic midplane. If we considered
NGC~891 with its parameters from \citet{rand90}, \citet{shaw89}, \citet{karachentsev13} 
as a part of our sample, it would fall into the high $M_s$ bin, high 
$L_{H\alpha-R_{eff}(r)}$ bin, and intermediate bin by $sSFR$. Its altitudes
z would span up to $(3\div 4) \, z_0$.
For comparison, parameters of the Milky Way from \citet{robitaille10}, \citet{licquia15}, \citet{mcmillan17}
would place it to the high $M_s$ bin, high $L_{H\alpha-R_{eff}(r)}$ bin,
and intermediate $sSFR$ bin — similar to NGC~891. 

We conclude that the assumption that eDIG on the BPT diagrams at any altitude 
in NGC~891 can be described only by photoionization models made by \citet{floresfajardo11}
is in agreement with our Figures 3–5. 
Therefore, the assumption about the OB-stars and HOLMES as the main sources of the DIG 
ionization based on a single galaxy, NGC~891, works well for the majority of other galaxies
and for a wide range of galactic altitudes. Nevertheless, it does not take into account 
possible contribution from shocks, which reveals itself at the highest altitudes, in the most
massive galaxies, and especially in galaxies with the lowest $sSFR$. Indeed, eDIG in our
galaxies can be described by photionization grids only, and we do not have to take  
possible contribution of shocks into account. 

In a general case (Figure 6), we observe increasing contribution of HOLMES and 
decreasing contribution of OB-stars to the gas ionization with the galactic altitude.
At low altitudes OB-stars contribute 3–5 times more to the ionization flux than
HOLMES. At the highest altitudes, in a contrary, HOLMES contribute 2–3 times
more to the gas ionization than OB-stars. The ionization parameter decreases 
systematically with the altitude. When the additional binning by the galactic 
parameters is applied, the trends described above are kept, qualitatively. 

When the sample is additionally binned by stellar mass $M_s$ (Figures 3 and 7), at any altitude
the OB-flux contribution decreases when the mass increases. In the least massive
galaxies the ionization parameter drops significantly with the altitude. 
The ionization parameter also increases with the mass at high galactic altitudes. 

When the \Ha\ luminosity $L_{H\alpha-R_{eff}(r)}$ is used for the binning (Figures 4 and 8),
at low galactic altitudes the OB-flux contribution increases with $L_{H\alpha-R_{eff}(r)}$, but this trend 
blurs at high altitudes. 

In the case of $sSFR$ binning (Figures 5 and 9) we see the most clear difference between the two main
sources of the gas ionization. At any galactic altitude increasing $sSFR$ means increasing
contribution of OB-stars to the ionizing flux. Moreover, in the high $sSFR$ galaxies OB-stars
contribute more than HOLMES at almost all galactic altitudes. Also the highest $sSFR$ galaxies 
demonstrate the most significant drop of the ionization parameter with altitude. 
In the low $sSFR$ galaxies the OB-stars contribution almost at all galactic altitudes
is less than that of HOLMES. This contribution varies with the altitude slowly,
as well as the ionization parameter does. 

This study is a logical extension of the work by \citet{jones17}, which uses much smaller
sample of galaxies, and we find that our conclusions agree with those by \citet{jones17}.
Thus, we observe systematic growth of forbidden emission lines with respect to the 
Balmer lines from the gas with the altitude increasing. Also, we confirm the guess
by Jones et al. that eDIG properties depend on some integral properties
of the galaxies, e.\,g. on stellar mass and specific star formation rate. 

We also notice that our results are in agreement with works by
\citet{belfiore16}, \citet{zhang17} that conclude that HOLMES are important for 
the DIG ionization. Unfortunately, direct comparison between our works 
is difficult because the galaxies considered by \citet{belfiore16}, \citet{zhang17}
allow them to study the DIG in the galactic midplanes only. 
The necessity of taking HOLMES into account for the description of 
the emission lines ratios in galactic gas is a significant addition
to the previous assumption that OB-stars can explain the ratios in the majority 
of galaxies made by \citet{haffner09}.

Promising conclusions were made in a study by \citet{lacerda18} based
on panoramic spectroscopy results from the survey CALIFA.
\citet{lacerda18} proposed to use the equivalent width (EW) of 
H$\alpha$ emission as an indicator of the gas ionization regime,
i.\,e. to distinguish the gas in DIG from HII regions. 
A humble sample of edge-on galaxies demonstrated by \citet{lacerda18}
gives a guess that the extraplanar gas is in the DIG ionization
regime, too. We note that our approach to the gas analysis makes
a reliable estimation of EWs difficult: while the flux from
emission lines can be reliably measured even at the highest altitudes, 
the stellar continuum is not well detected very far away from 
the galactic midplane. We take the EW analysis out of the frames of our work. 

At the end, we would like to discuss limitations of our approach to the 
gas state and ionization sources analysis at high galactic altitudes. Although
the spectra stacking procedure helps us increase the range where the eDIG 
can be studied, the stacked spectra are contributed by galaxies of different
kinds.  The limited spatial resolution of MaNGA does not allow us to resolve 
individual HII regions in eDIG. We can only extrapolate our knowledge
of smooth density distribution of gas at high galactic altitudes
earned from studies of our Galaxy. We also note that we do not 
separate galaxies by dominating gas ionization mechanisms, 
e.\,g. by the presence of active galactic nuclei. Also, we do not 
consider any radial binning in this paper, which would help 
us check whether powerful central sources can contribute to the
gas ionization. We also acknowledge a non-optimal determination 
of the vertical scale height of the disks via the effective radii. 
In this case, a two-dimensional photometric decomposition would
allow us to determine the vertical disk scale more reliably and to eliminate 
inconsistencies for the case of galaxies with large bulges. We purposely 
prefer to estimate the vertical scale from the effective radius for this 
study in order to better compare results to the work by \citet{jones17}.
We postpone the possible adjustments to our analysis mentioned above to 
a future work. 

\pagebreak 
\section*{Conclusions} 
 
We present results of a study of the ionization sources at different altitudes 
in galaxies of various types. The key element of this work is using a large
sample of 239 true edge-on galaxies selected from a recent SDSS data release DR16,
which allows us to study emission lines at extremely high altitudes via a
spectra stacking procedure. 

We compare the derived emission line ratios with results of modeling available from 
the 3MdB database, which enables us to consider the gas ionization by a combination of
OB-stars and HOLMES, and also by shocks. We employ three BPT diagrams for
the comparison. 

We find that the model of gas ionization that takes into account two types of sources:
OB-stars and HOLMES, adequately describes observed distributions of emission line ratios
in galactic gas. Nevertheless, shock waves may be required for better description
of DIG at high altitudes, especially at $z > 4.5 \, z_0$ in galaxies with intermediate
and high stellar masses or with low specific star formation rates. Interaction of gas in galaxies with circumgalactic medium 
\citep[see e.\,g.][]{slavin93} can be a source of the shocks, which also affects the gas kinematics
at high galactic altitudes \citep{bizyaev22}.

We infer how the OB-stars contribution to the total ionizing flux and the ionization
parameter change with the altitude via the grid interpolation from the BPT diagrams
and find the following trends:
 
 \begin{itemize}
     \item In the galaxies of all types (Figure 6) the contribution of OB-stars decreases, 
     the contribution of HOLMES increases, and the ionization parameter decreases with the 
     galactic altitude. 

     \item The increasing of stellar mass (Figures 3 and 7) leads to the OB-contribution decreasing and HOLMES
     contribution increasing at any galactic altitudes. Also, the vertical gradient of the ionization parameter
     decreases in this case. In turn, at the highest galactic altitudes the stellar mass increasing 
     leads to the increasing of the ionization parameter. 
 
     \item At fixed low galactic altitudes the growth of the $H\alpha$ luminosity (Figures 4 and 8) leads to 
     the increasing of OB-stars contribution and the decreasing of that from HOLMES. This trend blurs 
     at higher altitudes. 

     \item The most prominent difference in the considered sources of the  gas ionization 
     among all binning cases is seen for the galaxies with different specific star formation rates (Figures 5 and 9).
     With the specific star formation rate increasing, the OB-stars contribution increases and the one from HOLMES decreases.
     The ionization parameter vertical gradient also increases in this case. Moreover,
     for the galaxies with active star formation the contribution of OB-stars exceeds that from HOLMES
     at almost all galactic altitudes. In the star formation passive galaxies the OB-stars contribution is 
     less than that from HOLMES at almost all galactic altitudes, and it does not change significantly with galactic
     altitude, as well as the ionization parameter. 

 \end{itemize}

\section*{Acknowledgements}
This is a preprint of the work accepted for publication in Astronomy Letters,
©, copyright 2023, belonging to the authors, 
see http://pleiades.online/.
This study was partly supported by the Russian Science Foundation via 
grant 22-12-00080, https://rscf.ru/project/22-12-00080/. The authors thank the anonymous referee for his constructive feedback that improved the paper. 

The study makes use of the SDSS-IV MaNGA data available from  http://www.sdss4.org/dr16/data\_access/. The SDSS-IV 
website is https://www.sdss4.org. Funding for the Sloan Digital Sky 
Survey IV has been provided by the 
Alfred P. Sloan Foundation, the U.S. 
Department of Energy Office of 
Science, and the Participating 
Institutions. SDSS-IV acknowledges support and 
resources from the Center for High 
Performance Computing  at the 
University of Utah.

SDSS-IV is managed by the 
Astrophysical Research Consortium 
for the Participating Institutions 
of the SDSS Collaboration including 
the Brazilian Participation Group, 
the Carnegie Institution for Science, 
Carnegie Mellon University, Center for 
Astrophysics | Harvard \& 
Smithsonian, the Chilean Participation 
Group, the French Participation Group, 
Instituto de Astrof\'isica de 
Canarias, The Johns Hopkins 
University, Kavli Institute for the 
Physics and Mathematics of the 
Universe (IPMU) / University of 
Tokyo, the Korean Participation Group, 
Lawrence Berkeley National Laboratory, 
Leibniz Institut f\"ur Astrophysik 
Potsdam (AIP),  Max-Planck-Institut 
f\"ur Astronomie (MPIA Heidelberg), 
Max-Planck-Institut f\"ur 
Astrophysik (MPA Garching), 
Max-Planck-Institut f\"ur 
Extraterrestrische Physik (MPE), 
National Astronomical Observatories of 
China, New Mexico State University, 
New York University, University of 
Notre Dame, Observat\'ario 
Nacional / MCTI, The Ohio State 
University, Pennsylvania State 
University, Shanghai 
Astronomical Observatory, United 
Kingdom Participation Group, 
Universidad Nacional Aut\'onoma 
de M\'exico, University of Arizona, 
University of Colorado Boulder, 
University of Oxford, University of 
Portsmouth, University of Utah, 
University of Virginia, University 
of Washington, University of 
Wisconsin, Vanderbilt University, 
and Yale University.

{}

\end{document}